\documentstyle[a4wide,epsf,11pt,titlepage]{article}

\newcounter{nref}
\newcommand{\bbib}{%
  \renewcommand{\refname}{\large\bf References}%
  \begin{thebibliography}{99}%
  \setcounter{enumiv}{\thenref}}
\newcommand{\ebib}{%
  \end{thebibliography}%
  \setcounter{nref}{\arabic{enumiv}}}
\newcommand{\head}[3]{%
  \setcounter{nref}{0}%
  \thispagestyle{empty}%
  \section*{\LARGE\bf #1}%
  \stepcounter{section}%
  \addcontentsline{toc}{section}{#1}%
  \large\itshape%
  #2\\\vspace{0.1pt}\\%
  #3%
  \normalsize\upshape%
  \bigskip}

\begin{document}
\head{Cherenkov radiation by neutrinos}
     {Ara N.~Ioannisian$^{1,2}$, Georg G.~Raffelt$^2$}
     {$1$ Yerevan Physics Institute, Yerevan 375036, Armenia \\
      $2$ Max-Planck-Institut f\"ur Physik (Werner-Heisenberg-Institut), 
F\"ohringer Ring 6, 80805 M\"unchen, Germany}
\subsection*{Abstract}
We discuss the Cherenkov process $\nu\to\nu\gamma$ in the presence
of a homogeneous magnetic field. The neutrinos are taken to be
massless with only standard-model couplings.  The magnetic field
fulfills the dual purpose of inducing an effective neutrino-photon
vertex and of modifying the photon dispersion relation such that the
Cherenkov condition $\omega<|{\bf k}|$ is fulfilled.  
For a field strength $B_{\rm
crit}=m_e^2/e=4.41\times10^{13}~{\rm Gauss}$ and for $E=2m_e$ the
Cherenkov rate is about $6\times10^{-11}~{\rm s}^{-1}$.
\vspace{0.5cm}

In many astrophysical environments the absorption, emission, or
scattering of neutrinos occurs in dense media or in the presence of
strong magnetic fields \cite{ara1}. Of particular conceptual
interest are those reactions which have no counterpart in vacuum,
notably the decay $\gamma\to\bar\nu\nu$ and the Cherenkov
process $\nu\to\nu\gamma$. These reactions do not occur in vacuum
because they are kinematically forbidden and because neutrinos do not
couple to photons. In the presence of a medium or $B$-field, neutrinos
acquire an effective coupling to photons by virtue of intermediate
charged particles. 
In addition, media or external fields modify the dispersion relations of
all particles so that phase space is opened for neutrino-photon
reactions of the type $1\to 2+3$.

If neutrinos are exactly massless as we will always assume, and if
medium-induced modifications of their dispersion relation can be
neglected, the Cherenkov decay $\nu\to\nu\gamma$ is kinematically
possible whenever the photon four momentum $k=(\omega,{\bf k})$ is
space-like, i.e.\ ${\bf k}^2-\omega^2>0$.
Often the dispersion relation is expressed by $|{\bf k}|=n\omega$ in
terms of the refractive index~$n$. In this language the Cherenkov decay
is kinematically possible whenever $n>1$. 

Around pulsars field strengths around the critical value $B_{\rm
crit}=m_e^2/e=4.41\times10^{13}~{\rm Gauss}$.
The Cherenkov condition is satisfied for significant ranges of photon
frequencies. In addition, the magnetic field itself causes an
effective $\nu$-$\gamma$-vertex by standard-model neutrino couplings
to virtual electrons and positrons. Therefore, we study the Cherenkov
effect entirely within the particle-physics standard model.

This process has been calculated earlier in
~\cite{ara2}. However, we do not agree with their
results. 

Our work is closely related to a recent series of papers ~\cite{ara3}
who studied the neutrino radiative decay $\nu\to\nu'\gamma$ in the
presence of magnetic fields.

Our work is also related to the process of photon splitting that may
occur in magnetic fields as discussed, for example, in
Refs.~\cite{ara4,ara5}.  

Photons couple to neutrinos by the amplitudes
shown in Figs.~1(a) and (b). 
We limit our discussion to field
strengths not very much larger than $B_{\rm crit}=m_e^2/e$. 
Therefore, we keep only electron in the loop.  
Moreover, we are
interested in neutrino energies very much smaller than the $W$- and
$Z$-boson masses, allowing us to use the limit of infinitely heavy
gauge bosons and thus an effective four-fermion interaction (Fig.~1(c)).
The matrix element has the form
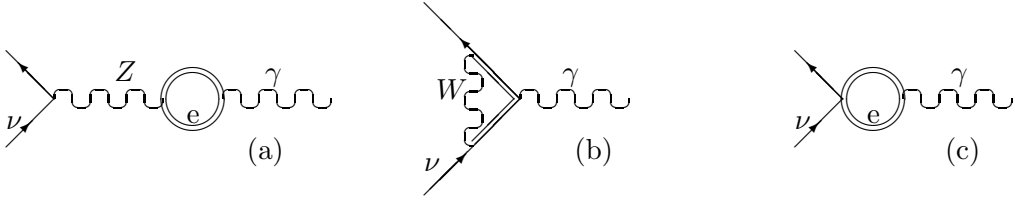
\begin{figure}
\centering\leavevmode
\vbox{
\unitlength=0.8mm
\begin{picture}(60,25)
\put(8,15){\line(-1,1){8}}
\put(8,15){\line(-1,-1){8}}
\put(0,7){\vector(1,1){4}}
\put(8,15){\vector(-1,1){6}}
\multiput(9.5,15)(6,0){3}{\oval(3,3)[t]}
\multiput(12.5,15)(6,0){3}{\oval(3,3)[b]}
\put(31,15){\circle{10}}
\put(31,15){\circle{9}}
\multiput(37.5,15)(6,0){3}{\oval(3,3)[t]}
\multiput(40.5,15)(6,0){3}{\oval(3,3)[b]}
\put(0,10){\shortstack{{}$\nu$}}
\put(18,18){\shortstack{{$Z$}}}
\put(43,18){\shortstack{{$\gamma$}}}
\put(30,11){\shortstack{{e}}}
\put(40,5){\shortstack{{(a)}}}
\end{picture}
\hspace{0.5cm}
\unitlength=0.8mm
\begin{picture}(60,32)
\put(16,15){\line(-1,1){7.5}}
\put(16,15){\line(-1,-1){7.5}}
\put(15,15){\line(-1,1){7}}
\put(15,15){\line(-1,-1){7}}
\put(16,15){\line(-1,1){16}}
\put(16,15){\line(-1,-1){16}}
\put(1,0){\vector(1,1){6}}
\put(16,15){\vector(-1,1){10}}
\multiput(17.5,15)(6,0){3}{\oval(3,3)[t]}
\multiput(20.5,15)(6,0){3}{\oval(3,3)[b]}
\multiput(8.2,8.7)(0,6){3}{\oval(3,3)[l]}
\multiput(8.2,11.7)(0,6){2}{\oval(3,3)[r]}
\put(2,15){\shortstack{{}$W$}}
\put(0,3){\shortstack{{}$\nu$}}
\put(23,18){\shortstack{{$\gamma$}}}
\put(25,5){\shortstack{{(b)}}}
\end{picture}
\unitlength=0.8mm
\begin{picture}(60,25)
\put(8,15){\line(-1,1){8}}
\put(8,15){\line(-1,-1){8}}
\put(0,7){\vector(1,1){4}}
\put(8,15){\vector(-1,1){6}}
\put(13,15){\circle{10}}
\put(13,15){\circle{9}}
\multiput(19.5,15)(6,0){3}{\oval(3,3)[t]}
\multiput(22.5,15)(6,0){3}{\oval(3,3)[b]}
\put(0,10){\shortstack{{}$\nu$}}
\put(26,18){\shortstack{{$\gamma$}}}
\put(12,11){\shortstack{{e}}}
\put(25,5){\shortstack{{(c)}}}
\end{picture}
}
\smallskip
\caption[...]{Neutrino-photon coupling in an external magnetic field.
The double line represents the electron propagator in the presence of
a $B$-field. 
(a)~$Z$-$A$-mixing. (b)~Penguin diagram (only for $\nu_e$).
(c)~Effective coupling in the limit of infinite gauge-boson masses.
\label{Fig1}}
\end{figure}
\begin{equation}
\label{m}
{\cal M}=-\frac{G_F}{\sqrt{2}\,e}Z\varepsilon_{\mu}
\bar{\nu}\gamma_{\nu}(1-\gamma_5)\nu\,
(g_V\Pi^{\mu \nu}-g_A\Pi_5^{\mu \nu}) ,
\end{equation} 
where $\varepsilon$ is the photon
polarization vector and $Z$ its wave-function renormalization
factor. For the physical circumstances of interest to us, the photon
refractive index will be very close to unity so that we will be able
to use the vacuum approximation $Z=1$.  
$g_V=2\sin^2\theta_W+\frac{1}{2}$ and 
$g_A=\frac{1}{2}$ for $\nu_e$, and
$g_V=2\sin^2\theta_W-\frac{1}{2}$ and
$g_A=-\frac{1}{2}$ for $\nu_{\mu,\tau}$.

Following Refs.~\cite{ara4,ara6,ara7,ara8} $\Pi^{\mu\nu}$ and
$\Pi_5^{\mu\nu}$ are
\begin{eqnarray}
\Pi^{\mu\nu}(k) &=& \frac{e^3B}{(4\pi)^2}
\Bigl[(g^{\mu \nu}k^2-k^{\mu}k^{\nu})N_0
-\,(g^{\mu \nu}_{\|}k^2_{\|}-k_{\|}^{\mu}k^{\nu}_{\|})N_{\|}+
(g^{\mu\nu}_{\bot}k^2_{\bot}-k^{\mu}_{\bot}k^{\nu}_{\bot})N_{\bot}
\Bigr], \nonumber\\
\Pi_5^{\mu \nu}(k) &=& \frac{e^3}{(4\pi)^2m_e^2}
\Bigl\{-C_\|\,k_{\|}^{\nu}(\widetilde{F} k)^{\mu}\
+ \ C_\bot\,\Bigl[k_{\bot}^{\nu}(k\widetilde{F})^{\mu}
+k_{\bot}^{\mu}(k\widetilde{F})^{\nu}-
k_{\bot}^2\widetilde{F}^{\mu \nu}\Bigr]\Bigr\},
\end{eqnarray} 
here  $\widetilde{F}^{\mu \nu}=
\frac{1}{2}\epsilon^{\mu \nu \rho \sigma}F_{\rho
\sigma}$, where $F_{12}=-F_{21}=B$. The $\|$ and $\bot$ decomposition of
the metric is
$g_\|={\rm diag}(-,0,0,+)$ and
$g_\bot=g-g_\|={\rm diag}(0,+,+,0)$. $k$ is the four
momentum of the photon.
$N_0$, $N_{\bot}$,$N_{\|}$, $C_\bot$ and $C_\|$ are
functions on $B$,$k^2_{\|}$ and $k^2_{\bot}$.
They  are real for $\omega<2m_e$, i.e.\ below the pair-production
threshold.  

The four-momenta conservation constrains the photon emission angle to have
the
value
\begin{equation}\label{emissionangle}
\cos \theta = \frac{1}{n} \
\left[1+(n^2-1)\frac{\omega}{2E}\right],
\end{equation}
where $\theta$ is the angle between the emitted photon and incoming
neutrino.
It turns out that for all situations of practical interest we have
$|n-1|\ll 1$ ~\cite{ara4,ara9}.
This reveals that the outgoing photon
propagates parallel to the original neutrino direction.

It is easy to see that the parity-conserving part of the effective vertex
($\Pi^{\mu \nu}$) is proportional to the small parameter
$(n-1)^2 \ll 1$ and the parity-violating part ($\Pi_5^{\mu \nu}$) is {\it
not\/}.
It is interesting to compare this finding with the standard 
plasma decay process $\gamma\to\bar\nu\nu$ which is dominated by the
$\Pi^{\mu \nu}$. Therefore, in the approximation
$\sin^2\theta_W=\frac{1}{4}$ only the electron
flavor contributes to plasmon decay. Here the Cherenkov rate is equal for 
(anti)neutrinos of all flavors.

We consider at first neutrino energies below the pair-production
threshold $E<2m_e$. For $\omega<2m_e$ the photon refractive
index always obeys the Cherenkov condition $n>1$ ~\cite{ara4,ara9}.
Further, it turns out that in the range
$0<\omega< 2m_e$ $C_\|$,$C_\perp$ depend only
weakly on $\omega$ so that it is well approximated by its value at
$\omega=0$.  
For neutrinos which propagate perpendicular to the magnetic
field, a Cherenkov emission rate can be written in the form
\begin{eqnarray}\label{finalresult}
\Gamma\ \approx \ \frac{4\alpha G_F^2E^5}{135(4\pi)^4}\,
\left(\frac{B}{B_{\rm crit}}\right)^2 h(B)\
= \ 
2.0\times10^{-9}~{\rm s}^{-1}~\left(\frac{E}{2m_e}\right)^5
\left(\frac{B}{B_{\rm crit}}\right)^2 h(B),
\end{eqnarray}
where 
\begin{equation}
h(B)= 
\cases{(4/25)\,(B/B_{\rm crit})^4&for $B\ll B_{\rm crit}$,\cr
1&for $B\gg B_{\rm crit}$.\cr}
\end{equation}
Turning next to the case $E>2m_e$ we note that in the presence of a
magnetic field the electron and positron wavefunctions are Landau
states so that the process $\nu\to\nu e^+e^-$ becomes kinematically
allowed. Therefore, neutrinos with such large energies will
lose energy primarily by pair production rather than by Cherenkov
radiation (for recent calculations see ~\cite{ara10}).

The
strongest magnetic fields known in nature are near pulsars. However,
they have a spatial extent of only tens of kilometers. Therefore, even
if the field strength is as large as the critical one, most neutrinos
escaping from the pulsar or passing through its magnetosphere will not
emit Cherenkov photons. Thus, the magnetosphere of a pulsar is quite
transparent to neutrinos as one might have expected.
\subsection*{Acknowledgments}
It is pleasure to thanks the organizers of the Neutrino Workshop at the
Ringberg Castle for organizing a very interesting and enjoyable
workshop. 

\end{document}